\documentclass{optica-article}

\journal{opticajournal} 

\articletype{Research Article}

\usepackage{placeins}
\usepackage{lineno}

\begin{document}

\title{Ultra-fast and accurate multimode waveguide design based on dataset-based eigenmode expansion method}

\author{Jaesung Song,\authormark{1} and Young-Ik Sohn\authormark{1,*}}

\address{\authormark{1}School of Electrical Engineering, Korea Advanced Institute of Science and Technology (KAIST), Daejeon 34141, Republic of Korea\\
}

\email{\authormark{*}youngik.sohn@kaist.ac.kr} 


\begin{abstract*} 
We propose a dataset-based photonic simulation framework for multimode waveguide design, enabling ultra-fast simulations with high accuracy. Compared to conventional approaches, our method offers two to three orders of magnitude speed-up in complex multimode waveguide designs. Based on this approach, we demonstrate a silicon multimode waveguide bend with an effective radius of $30\,\mu\mathrm{m}$ under one second, with accuracy validated against commercial 3D finite-difference time-domain method. We further explore its utility for device optimization by designing a $20\,\mu \mathrm{m}$-radius, arbitrary power splitting ratio bends through thousands of optimization iterations, completed in just 68 minutes on a standard desktop CPU. This framework enables the rapid design and engineering of large-scale multimode photonic devices, making computationally intensive simulations more accessible to many photonic circuit designers.

\end{abstract*}

\section{Introduction}
The photonic integrated circuit (PIC) has recently become a key industry-scale technology as hardware production is more mature and new applications are found \cite{photonics_advantages_Shaker_2023}. For example, it is increasingly recognized as a promising platform for sensing applications such as LiDAR as well as photonic computational processing. The latter, often referred to as `xPU' in specialized research communities, is especially receiving a great amount of attention because of its potential for high energy efficiency and parallel processing capabilities \cite{Roadmapping_photonics_Shekhar_2024}. In addition to areas already deployed in the market, more applications are actively being discovered, such as quantum technologies. In the case of silicon photonics, the most dominant PIC technology at the moment, its compatibility with mature CMOS processes enables scalable fabrication and integration, making it a compelling candidate for universal quantum computing \cite{PsiQuantum_2025, Llewellyn_2020}.

Although PIC technology has rapidly progressed, it is still not fully utilizing the potential of photonic hardware. For example, current single-mode waveguide circuits are not very efficient in using the space resource of the chip. This inefficiency is fundamentally due to the nature of mainstream PIC technology, in which light typically passes through a given device only once in a single spatial mode. Currently, such single-pass single-mode waveguide operation is dominant in commercial applications because of its reliability and predictability \cite{Stirling_2022}. However, it is not an ideal mode of operation in terms of resource usage, since there is a room for applying various multiplexing schemes, as was done for optical communication technologies.

To address this limitation, multimode waveguide technology has been investigated in the past decade and is currently becoming more practical as a result of such efforts \cite{Chack_2020, Li_2018, Jiang_2018, Zhu_2024, Xu_2021, mm_arb_angle_Lan_2025}. Therefore, the wide adoption of multimode waveguides in PICs is expected to enhance spatial efficiency by almost an order of magnitude compared to the current single-pass single-mode approach\cite{Hong_2025}. 

At the same time, multimode technology is also essential even when it is operated with a single mode only because of the very low propagation loss. Because multimode waveguides typically have wider widths than their single-mode counterparts, the optical mode exhibits reduced overlap with rough sidewalls, resulting in decreased scattering loss \cite{Melati_2014, Ji_2021, Cui_2024}. Minimizing propagation loss is becoming more important as circuits become larger and more complex \cite{Bogaerts_2020}. Furthermore, ultra-low-loss performance is especially critical in many quantum applications such as quantum computing \cite{PsiQuantum_2025}.

Although previously considered undesirable due to intermodal crosstalk, with proper engineering, multimode waveguides are very useful not only for low-loss operation but also for many other purposes. Multimode waveguide technology has become increasingly popular recently due to additional multiplexing advantages that are not allowed for single-mode waveguides. In particular, they facilitate mode division (de)multiplexing, either as an alternative or in combination with frequency division multiplexing, thus optimizing the transmission capacity of data \cite{Dai_2013, Li_2018, Chack_2020}. The multimode platform has also enabled innovations such as waveguide recycling, which reduces the footprint of true-time delay lines \cite{light_recycling_Koh2022, Hong_2025} and improves the efficiency of phase shifters in phased array systems \cite{recycling_OPA_Miller_2020}. Recent studies have further explored the use of multimode waveguides in quantum applications, including quantum interference\cite{Mohanty_2017} and the formation of photonic molecules\cite{Tao_2024}, unlocking new possibilities for quantum phenomena and novel photonic interactions. 

However, a key challenge remains before multimode technology can be considered reliable: the precise control of multiple spatial modes. Poorly engineered structures can cause unwanted intermodal crosstalk, degrading signal quality \cite{Gabrielli_2012}. Therefore, a robust design and optimization process is essential to control mode interactions and suppress unwanted mode excitations, ensuring stable device performance \cite{Liao_2024, Li_2020}.

Optimizing photonic structures often requires conducting a few dozen to several thousand simulations \cite{Xu_2025, Liao_2024, guo2023, Michaels_2018}. Therefore, even a modest reduction in simulation time and computational cost can significantly impact overall optimization efforts. Although established methods, such as the finite difference time domain (FDTD) method \cite{Teixeira_2023} and the eigenmode expansion (EME) method \cite{Bienstman_2001} provide accurate results, they are computationally intensive and time consuming \cite{Yu_2024, guo2023}. Accordingly, improvements in simulation efficiency are critical for accelerating the design and optimization of next-generation photonic devices.

In this paper, we demonstrate a dataset-based simulation method for photonic structures that significantly reduces both the simulation time and computational cost. We begin with a brief review of the conventional EME method, which serves as the foundation for our approach. Next, we introduce the dataset-based method and its overall simulation framework, examining its accuracy in relation to dataset resolution. We then demonstrate how a verified dataset can substantially accelerate simulations. Finally, we explore various optimization applications enabled by this approach and discuss its potential for a broader implementation in photonic design.

\section{Simulation Principle}
\subsection{Conventional EME method}
\begin{figure*}[ht]
\centering
\includegraphics[width=0.9\textwidth]{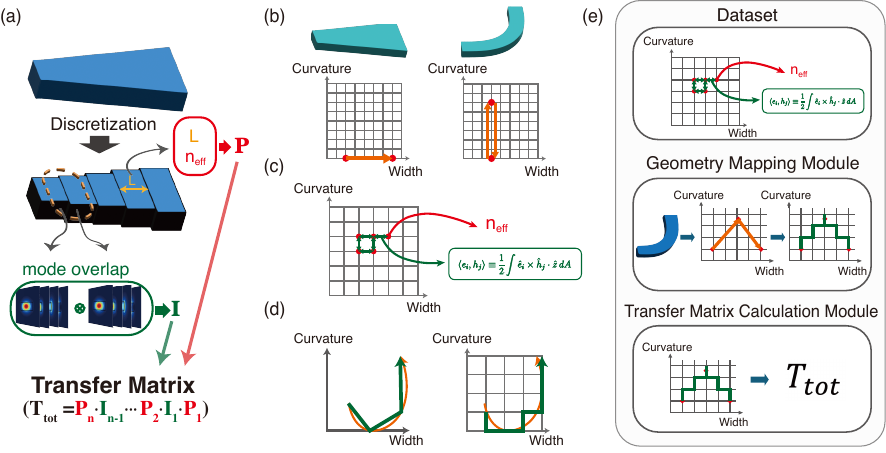}
\caption{Principle of dataset-based photonic simulation. (a) Conventional EME method: the structure is discretized, and phase evolution matrices $P$ and interface matrices $I$ are computed using $n_{eff}$ and mode overlaps, respectively, to form the total transfer matrix $T_{tot}$.
(b) Geometric representation of taper and Euler bend structures as trajectories in width–curvature space.
(c) Discretized dataset: effective indices are stored at grid points, and mode overlaps at grid edges.
(d) Structure discretization in the conventional EME method (left) and the dataset-based EME method (right).
(e) Dataset-based simulation framework with three modules: dataset, geometry mapping, and transfer matrix calculation.}
\label{fig:simulation_principle}
\end{figure*}

The conventional EME or mode-matching method is a widely used technique to calculate the transfer matrix or, equivalently, the scattering matrix of a given photonic structure \cite{Bienstman_2001}. A schematic representation of the EME simulation method is provided in Fig. \ref{fig:simulation_principle}(a). The method begins by partitioning the target structure into multiple slices perpendicular to the light propagation axis, thereby segmenting the structure. Each segment is then approximated as having a uniform cross-section, simplifying the modeling of waveguide properties.

Subsequently, an eigenvalue problem is solved for each uniform segment to determine the eigenmode field profiles and their corresponding effective indices. Following this, the mode overlap matrix, $O$, is computed, where the element in the $i$-th row and $j$-th column, $O_{ij}$ is given by the mode overlap integral: 

\begin{equation}
O_{ij} \equiv \frac{1}{2} \int \hat{e}_i \times \hat{h}_j \cdot \hat{z} \, dA,
\label{eq:mode_overlap}
\end{equation}
where $\hat{e}_i$ and $\hat{h}_j$ denote the normalized $i$-th mode electric field (E-field) and $j$-th mode magnetic field (H-field), respectively, and $\hat{z}$ denotes the unit vector along the propagation axis.

Energy transfer between eigenmodes does not occur within a uniform cross-sectional segment, allowing the uniform structure to be represented by a phase evolution matrix, $P$. The phase evolution is given by $e^{i\beta z}$ where $\beta = 2\pi n_{\text{eff}} / \lambda$ is the propagation constant, $n_{\text{eff}}$ is the effective refractive index of the mode, and $z$ represents the propagation distance. Thus, the phase evolution matrix is determined by the effective indices $n_{eff}$ and the segment length $L$. At the interfaces between discretized segments, the mode-matching method is employed to compute the transfer matrix by enforcing Maxwell’s boundary conditions, which ensure the continuity of transverse fields. Based on these conditions, the interface transfer matrix $I$ is obtained primarily through basic matrix operations, including addition, subtraction, and inversion of the mode overlap matrix. Finally, the total transfer matrix, $T_{tot}$ of the structure is obtained by sequentially multiplying the phase propagation matrix $P$ with the interface transfer matrix $I$.

\subsection{Geometric Parameter Space}
Photonic structures, including waveguide structures, can be described as functions of geometric parameters such as width and curvature along the propagation axis \cite{Tambasco_2023}. For instance, a taper can be defined by a width parameter function, whereas a bend with a constant width can be characterized by a curvature parameter function.

Building on this idea, an abstract geometric parameter space can be conceptualized where each axis corresponds to an independent geometric parameter. In this space, any photonic structure can be represented as a trajectory, provided that an appropriate set of geometric parameters is chosen. This representation describes how the geometry of a structure changes by focusing on the evolution of the trajectory path within the parameter space, rather than how those parameters vary along the length of the physical propagation axis, $z$. In this abstract space, the structure is no longer represented as a function of $z$, but rather as a sequence of parameter values such as multiple pairs of width and curvature. Fig. \ref{fig:simulation_principle}(b) illustrates this concept using examples of taper and bend structures.

\subsection{Discretized Dataset}
In conventional photonic device optimization methods based on the iteration approach, even slight variations in the structure previously calculated require recomputing the entire calculation from the beginning. For instance, in FDTD simulations, even a minor geometric change modifies the way fields propagate, scatter, and interact with boundaries. Since the field distribution at a given moment depends on the evolution over the past, this small change makes the entire simulation invalid and necessitates a full restart of the simulation to accurately capture the updated behavior.

Similarly, when using the EME method, as illustrated on the left side of Fig. \ref{fig:simulation_principle}(d), the structure is discretized in geometric parameter space.  To simulate such discretized structures, one must compute the eigenmodes of cross-sections at sampled points along the structure trajectory and calculate the overlap matrices between eigenmodes at adjacent sampled points. Because the typical trajectory lies in a continuous geometric parameter space, even minor structural variations alter many sampling points which correspond to eigenmodes and overlap matrices used in the transfer matrix calculations. Consequently, conventional EME requires recomputing all the steps for each simulation, resulting in significant computational redundancy.

To address this inefficiency, we introduce a dataset based on a discretized geometric parameter space. Fig. \ref{fig:simulation_principle}(c) illustrates this discretized dataset. Each geometric parameter is discretized, forming a grid in the geometric parameter space. The intersection points in this grid correspond to specific cross-sectional geometries. At each grid intersection, we store the effective indices of the eigenmodes associated with the corresponding cross-section. Additionally, at the edges connecting adjacent intersection points, we store the overlap matrices, where the field profiles are defined at each intersection point.

\subsection{Dataset-based Simulation}
Any structure defined by geometric functions can be mapped approximately to this discretized geometric parameter space. An example of this mapping is shown in the right panel of Fig. \ref{fig:simulation_principle}(d). In this representation, the structural trajectory becomes a piecewise path composed of horizontal and vertical segments aligned with the grid axes. This effect, referred to here as grid-induced structure distortion, occurs because the mapped path includes grid points that do not lie on the original structure's path. As the resolution of the grid increases, the discretized trajectory more closely approximates the true geometry, reducing the associated error.

Once a properly constructed dataset is available, the transfer matrix can be calculated solely using the effective indices and overlap matrices stored following the EME method. This approach reduces the simulation to a simple product of precomputed two-dimensional matrices, thereby eliminating the need to compute eigenmodes and overlap matrices repeatedly on demand, which dominate the simulation workload.

Our dataset-based simulation framework is illustrated in Fig. \ref{fig:simulation_principle}(e). It consists of three main components: the dataset, the geometry mapping module, and the transfer matrix calculation module. The dataset is constructed on the basis of discretized geometric parameter space, where the choice of geometric parameters depends on the structure being simulated. The dimension of the parameter space is determined by the number of controllable parameters in the cross-section. For instance, for a single waveguide structure made of an isotropic material such as rectangular waveguides in silicon photonics, width and curvature give full information of the cross-section. The geometry mapping module approximates the given photonic structure to discretized geometries which are placed on the dataset's grid points in parameter space. Finally, the transfer matrix calculation module computes the transfer matrix for the discretized structure using the precomputed data stored in the dataset.

An important feature of this framework is that dataset generation relies solely on localized computations — eigenmodes at each grid point and overlap integrals between adjacent points — making it inherently parallelizable. Multiple computing resources can independently generate partial datasets, which are combined afterwards to form the complete dataset. This architecture not only eliminates redundant simulation work across multiple structures but also enables scalable and efficient dataset generation.

\section{Balancing Dataset Resolution and Performance}
The accuracy of approximating the original structure through discretization strongly depends on the grid spacing used in the dataset. However, excessively high resolution increases the computational cost of dataset generation. Therefore, selecting the appropriate resolution for the grid requires balancing accuracy with computational efficiency. In this paper, we focus on rectangular waveguide structures fabricated from silicon, where the only varying geometric parameters are width and curvature. The two-dimensional parameter space limits the size of the dataset, and hence it is possible to achieve high accuracy even with a standard desktop CPU.

\begin{figure}[ht]
\centering
\includegraphics[width=0.5\linewidth]{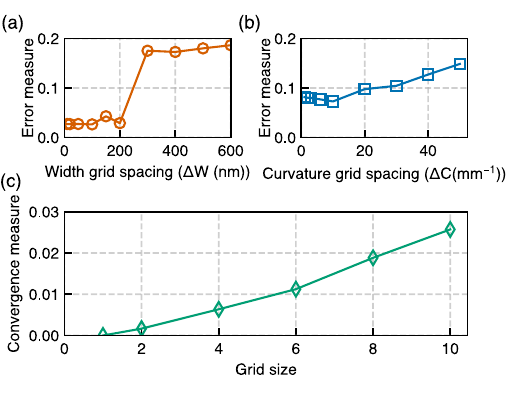}
\caption{Accuracy evaluation of dataset resolution. (a), (b) Error measure for test structures as a function of grid spacing for (a) width ($\Delta W$) and (b) curvature ($\Delta C$). (c) Convergence measure quantifying deviation from the finest-grid simulation (grid size $i$=1) as a function of grid size for jointly varying parameters.}
\label{fig:dataset_convergence}
\end{figure}

Discretization errors in the dataset-based EME method can be classified into two categories: grid-induced structure distortion and intrinsic discretization error of the original EME approach. As described in the previous section, grid-induced structure distortion refers to deviations in the structural trajectory introduced by snapping to the dataset grid, resulting in a stair-step or zigzag path in geometric parameter space. In contrast, intrinsic discretization error arises from an insufficient number of discretized segments — even when all discretized points lie on the actual structural trajectory — due to the limitations of coarse sampling in accurately capturing continuous geometric variation.

To address the two types of errors independently, we adopted a two-stage evaluation process to select the resolution of the dataset: (1) testing structures with variation in a single parameter and (2) testing structures with simultaneous variation in multiple parameters. In the first stage, the structural trajectory lies parallel to a single axis in the geometric parameter space. As a result, grid-induced structure distortion does not occur, allowing us to isolate and evaluate the intrinsic discretization error. In the second stage, multi-parameter structures produce trajectories that traverse multiple grid directions, introducing grid-induced structure distortion. This stage enables the assessment of errors specifically caused by the snapping error only.

For the first stage, we estimate an appropriate resolution for each individual parameter through convergence testing. Each test structure was designed such that only one parameter, width or curvature, was affected, while the other remained fixed. The accuracy was evaluated by comparing the simulation results from the dataset-based method with the reference FDTD simulations. The width test structures included linear tapers and tapers with arbitrarily varying width profiles at different lengths. The curvature test structures consisted of Euler bends and 90 degree bends with arbitrary curvature variations, all maintaining a constant waveguide width. Further details of the test structures are provided in Appendix \ref{app:test_structures}.

To quantify the accuracy of datasets of different grid spacings, we defined the following error metric:
\begin{equation}
\text{Error measure} = \frac{\left| |S_{\text{ref},1}|^2 - |S_{\text{test},1}|^2 \right| + \left| |S_{\text{ref},2}|^2 - |S_{\text{test},2}|^2 \right|}{2},
\label{eq:error_measure}
\end{equation}
where $S_{\text{ref},i}$ and $S_{\text{test},i}$ denote the scattering matrix elements from the input mode to the $i$-th dominant transmitted mode, obtained from the reference FDTD simulation and the dataset-based EME simulation, respectively. Figs. \ref{fig:dataset_convergence}(a),(b) present this error measure as a function of parameter grid spacing for width ($w$) and curvature ($C$), respectively. We observed improved accuracy as the grid spacing decreased, eventually converging to a residual error primarily attributable to inherent limitations of the EME method. Such inherent limitations include the finite number of modes used in calculation and the limited ability to accurately model the abrupt cross-section variations in geometry. The coarsest grid spacings that still achieved convergence, as identified in this analysis, were $\Delta w = 100\ \text{nm}$ for width and $\Delta C = 10\ \text{mm}^{-1}$ for curvature. These values represent the resolution limits beyond which the intrinsic discretization error is suppressed as much as possible.

Having established parameter resolutions sufficient to suppress intrinsic error, we proceeded to assess errors arising from grid-induced structure distortion in the second stage. We utilize pre-defined test structures in which multiple geometric parameters vary simultaneously. These structures included tapered partial Euler bends and 90-degree bends with arbitrarily varying width and curvature profiles across different radii. Further details of the test structures are provided in Appendix \ref{app:test_structures}.

We adopted the previously determined converged resolutions as baseline grid spacings by fixing the resolution ratio between the two parameters. To do so, we introduced a scaling factor, defining the grid size $m$ such that $\Delta w = m \times 10\ \text{nm}$ and $\Delta C = m \times 1\ \text{mm}^{-1}$. Therefore, the baseline grid found from the previous step corresponds to $m = 10$. Using the finest available grid size ($m = 1$) as a reference, we established a convergence metric for the second stage:
\begin{equation}
\text{Convergence measure} = \left| |S_{\text{ref},ij}|^2 - |S_{\text{test},ij}|^2 \right|,
\label{eq:convergence_measure}
\end{equation}
where $S_{\text{ref},ij}$ and $S_{\text{test},ij}$ correspond to scattering matrix elements connecting input mode $i$ to output mode $j$, obtained from the reference simulation (grid size $m=1$) and the test datasets at coarser resolutions, respectively. We considered the first four guided modes for both input and output ports. By definition, this convergence measure approaches zero as the resolution of the test dataset improves toward the reference dataset (grid size $m=1$). The results are shown in Fig. \ref{fig:dataset_convergence}(c), which plots the convergence measure as a function of grid size.

\begin{figure}[ht]
\centering
\includegraphics[width=0.5\linewidth]{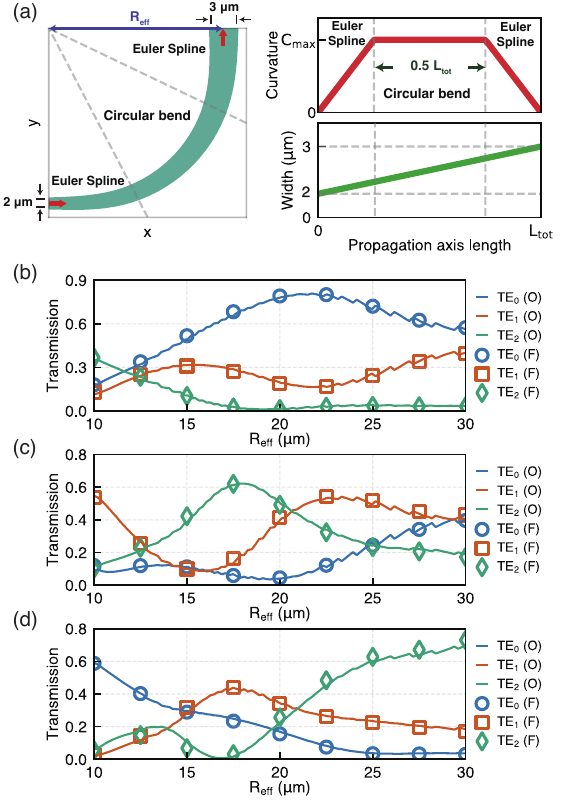}
\caption{Validation of dataset-based simulation using tapered partial Euler bends.
(a) Structure schematic with a linearly tapered width and a composite curvature profile consisting of Euler spline and circular bend segments.
(b)–(d) Transmission as a function of effective bend radius (\(R_{\mathrm{eff}}\)) for different input modes:
(b) \(\text{TE}_0\),
(c) \(\text{TE}_1\),
and (d) \(\text{TE}_2\).
Solid lines: dataset-based EME results; open markers: reference 3D FDTD results (mesh size: \(\lambda/15\)).
The strong agreement demonstrates the accuracy of the proposed method.}
\label{fig:simulation_result_comparison}
\end{figure}

The dataset generation time and size scale proportionally to the number of intersection points in the parameter grid. To manage the computational load, we employed up to three standard desktop computers to parallelize the dataset generation process. The dataset was generated at a wavelength of 1550 nm, covering widths from $1\,\mu\mathrm{m}$ to $3\,\mu\mathrm{m}$ and curvatures from $0\,\mathrm{mm}^{-1}$ (straight waveguide) to $330\,\mathrm{mm}^{-1}$ (corresponding to a bend radius of $3.03\,\mu\mathrm{m}$). The generation times for different grid resolutions $m$ were approximately 4 hours for $m=10$, 16 hours for $m=5$, 93 hours for $m=2$, and 370 hours for $m=1$. Based on this analysis, we selected $m=2$ as the optimal resolution, which ensures the dataset can be generated in one week while maintaining sufficient accuracy for the simulations presented in this study. The resulting dataset size under these conditions was approximately 350 MB only.

To illustrate the accuracy and robustness of our dataset-based simulation approach, we analyzed partial Euler bend structures that exhibit simultaneous variations in multiple parameters. Fig.~\ref{fig:simulation_result_comparison}(a) presents the top view of the partial Euler bend along with its width and curvature functions. The structures feature a linearly tapered width, increasing from $2 \mu \mathrm{m}$ at the input to $3\,\mu \mathrm{m}$ at the output, and a partial Euler spiral configuration having a spiral fraction of 0.5. Consequently, the curvature increases and decreases linearly in the first and last quarters of the bend, while remaining constant in the central region. Figs. \ref{fig:simulation_result_comparison}(b)–(d) compare the transmission results obtained from our dataset-based simulations (O) against the reference FDTD simulations (F) as a function of the effective bend radius for three distinct input modes. Although slight fluctuations due to discretization artifacts were observed in the dataset-based results, they closely matched the overall trends of the reference FDTD simulations. These discretization-induced fluctuations can be reduced further by employing datasets with higher grid resolutions.

\section{Simulation time}
Reducing simulation time and computational cost is the key advantage of our approach, particularly when optimizing large-scale photonic devices, which typically require many simulations. To illustrate the usefulness of our dataset-based simulation method, we compare its computational performance with conventional methods, namely 3D FDTD and conventional EME. Those methods are the primary 3D simulation techniques employed in photonic device design. Therefore, a comparison with 3D FDTD and conventional EME is sufficient to evaluate the performance of our approach. Table \ref{tab:Simulation_time} summarizes this comparison for a tapered partial Euler bend structure at various effective radii, which correspond directly to the dimensions of the simulated structure.

As shown in Table \ref{tab:Simulation_time}, the simulation time for FDTD increases significantly with larger structure sizes because its computational complexity scales directly with the simulation volume. In contrast, simulation times for conventional EME remain relatively constant as the number of discretized segments does not strongly depend on the structure size. However, despite this consistency, each conventional EME simulation still demands considerable computational resources due to repeated calculations of eigenmodes and overlap matrices.

\begin{table}[htbp]
\centering
\caption{\bf Comparison of Simulation Times for Different Methods$^{\textit{a}}$}
\begin{tabular}{c|ccc}
\hline
$R_{\text{eff}}$ & Ours$^{\textit{b}}$ & 3D FDTD$^{\textit{c}}$ & Conv. EME$^{\textit{d}}$ \\ 
\hline
$10\,\mu$m & 0.77 s   & 32 min & 11 min 22 s\\
$20\,\mu$m & 0.3 s & 2 hr 47 min   & 13 min 8 s\\
$30\,\mu$m & 0.22 s & 10 hr 25 min & 13 min 43 s\\
\hline
\end{tabular}
\label{tab:Simulation_time}
\vspace{0.5em}

\begin{minipage}{0.9\linewidth}
\footnotesize
\textit{a}~All simulations were performed on an Intel i9-9900K CPU. \\
\textit{b}~Average values for 37 runs.\\
\textit{c}~Mesh size: $\lambda/22$. \\
\textit{d}~50 sections; FDE domain size: $4\,\mu\text{m} \times 2\,\mu\text{m}$; mesh size: 10 nm (core), 20 nm (cladding).
\end{minipage}
\end{table}

Our proposed dataset-based EME method demonstrates a drastic improvement in computation efficiency. Remarkably, its simulation times are on the order of seconds or even subseconds, which is negligible compared to conventional methods. The simulation speed improvement holds for tapers, bends and the combination of the two with a single dataset. Moreover, since the simulation relies solely on the multiplication of precomputed matrices, the same advantage is expected to extend to more complex structures, such as coupled waveguides, once the corresponding dataset is prepared.

A notable observation is that the simulation time decreases slightly as the size of the structure increases, a somewhat counterintuitive result. This behavior occurs because larger bend structures typically traverse simpler or smaller regions in the geometric parameter space, thereby requiring fewer matrix computations. Consequently, our dataset-based approach provides substantial reductions in simulation time, especially as the structure dimensions grow. The subsecond and 10 min computation time differences may not be important for simulating a given structure; however, it can make a huge impact when optimizing devices, which typically needs tens or hundreds of iterations.

\section{Optimization Examples}
Optimization of photonic structures refers to a systematic search for structural designs that achieve the best optical performance for given purposes, such as minimal loss or maximal mode coupling. Each tunable parameter that affects the performance of the photonic structure is considered an optimization variable. These variables define a multidimensional optimization parameter space, where each point corresponds to a unique structural design. Each point in the optimization parameter space defines a unique trajectory in the geometric parameter space. Thus, optimization can be viewed as identifying the optimal point within this optimization parameter space. To determine the optimal design, simulations must be performed for the photonic structure represented by each candidate point in the optimization space.

\begin{figure*}[ht]
\centering
\includegraphics[width=\textwidth]{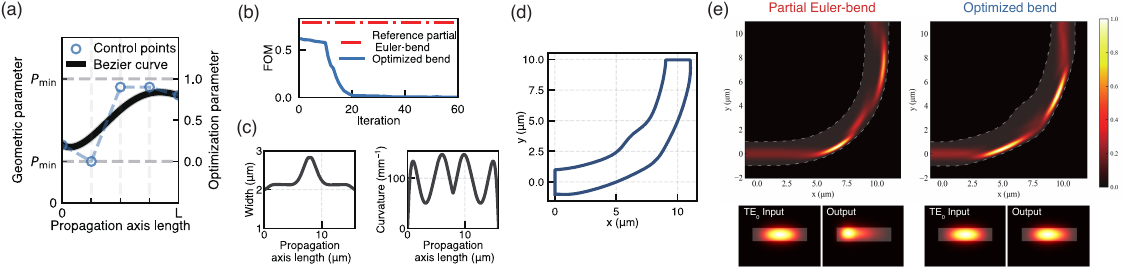}
\caption{Bend optimization procedure and results for an effective radius of $10\, \mu \mathrm{m}$ and input/output widths of $2\,\mu \mathrm{m}$.
(a) Parameterization of the geometric function using a Bézier curve. Control points (circles) define the shape of the curve.
(b) Evolution of the FOM during optimization. The red dash-dot line indicates the FOM of the reference partial Euler bend (spiral fraction =  0.5). FOM for an ideal mode-preserving bend is supposed to be zero.
(c) Optimized width and curvature as function of propagation axis length.
(d) Top-view lay out of the bend shape mapped from the optimized geometric functions.
(e) Top-view power-flow distributions and mode profiles at the input and output. The optimized bend shows improved mode preservation compared to the reference Euler bend.}
\label{fig:reff10um_optimization}
\end{figure*}

Optimization methods generally fall into two main categories: single-solution-based methods and population-based methods. Single-solution-based methods iteratively refine a single candidate solution by exploring neighboring regions locally within the parameter space \cite{GA_katoch2021}. A representative example is the adjoint method
\cite{Keraly_2013, Michaels_2018}, which efficiently computes gradients in the parameter space using only two FDTD simulations per iteration (one forward and one backward). However, despite its efficiency in gradient estimation, the adjoint method still involves computationally intensive FDTD simulations, making it particularly costly and time consuming when optimizing large-scale structures. Moreover, single-solution-based methods, including the adjoint approach, are prone to become trapped in local optima \cite{Michaels_2018}.

In contrast, population-based methods simultaneously evaluate multiple candidate solutions, increasing the likelihood of finding a global optimum \cite{mm_arb_angle_Lan_2025}. Examples of population-based methods include Particle Swarm Optimization (PSO) \cite{PSO_Gad_2022, mm_arb_angle_Lan_2025} and Genetic Algorithms (GA) \cite{mm_arb_angle_Lan_2025}. Despite their advantage in exploring the global parameter space, population-based methods typically require a substantial number of simulations per iteration, greatly increasing computational demands.

To overcome the limitations inherent to both methods, hybrid optimization approaches have been proposed. These hybrid methods first use a population-based approach to identify the region near the global optimum, after which the adjoint method is employed to efficiently fine-tune the solution \cite{Liao_2024, Liao_2024_2}.

Our dataset-based EME method can bring enormous advantage in all three categories, single-solution-based method, population-based method, and hybrid approach. This is because the calculation time for a given structure is improved by two to three orders of magnitude while the accuracy is barely compromised. To showcase its capability, we applied it to optimize several multimode bends using population-based algorithms, which typically require a large number of simulations. 

For these optimizations, we parameterize the geometric parameter function as Bézier curve which provides smoothness and is easily controlled through a set of control points. Fig. \ref{fig:reff10um_optimization}(a) illustrates this parameterization where normalized optimization parameters serve as control points that define a Bézier-shaped geometric function. Based on the inherent smoothness of the Bézier curves, we can limit our design space to structures with very low loss.

As a first demonstration, we optimized a multimode waveguide bend structure with an effective radius of $10\,\mu \mathrm{m}$ and input/output widths of $2\,\mu \mathrm{m}$. The optimization aimed to preserve the fundamental $\text{TE}_0$ mode at the bend output. We defined the objective function as a figure of merit (FOM):
\begin{equation}
\text{FOM} = \big|1 - |S_{11}|^2| - penalty
\label{eq:FOM}
\end{equation}
where $S_{11}$ is the scattering matrix element representing transmission from the fundamental input mode back into the same mode at the output. The penalty term is introduced to discourage solutions with an excessively small curvature and is defined as: $\text{penalty} = \min\left[0, \frac{325 - \max(\text{curvature})}{10} \right]$. This imposes a penalty when the maximum curvature exceeds \(325\,\text{mm}^{-1}\), guiding optimization to remain within the curvature range of the dataset. The optimization was performed using Particle Swarm Optimization (PSO), a widely adopted population-based method in integrated photonics design. A population of 100 particles was evolved over 60 iterations, totaling 6,000 individual simulations. Conducted on an Intel i9-9900K CPU, the entire optimization required approximately 137 minutes. Fig. \ref{fig:reff10um_optimization}(b) shows the optimization FOM history, clearly illustrating convergence, while Fig. \ref{fig:reff10um_optimization}(c) shows the optimized geometric functions. The resulting bend shape, mapped from these functions, is illustrated in Fig. \ref{fig:reff10um_optimization}(d). 

We validated the performance of the optimized structure using 3D FDTD simulations with mesh size $\lambda/22$. Fig. \ref{fig:reff10um_optimization}(e) shows the FDTD results, comparing the power flow profiles for the optimized bend and a conventional partial Euler bend, confirming the superior mode-preserving performance of the optimized structure.

\begin{figure}[ht]
\centering
\includegraphics[width=0.5\linewidth]{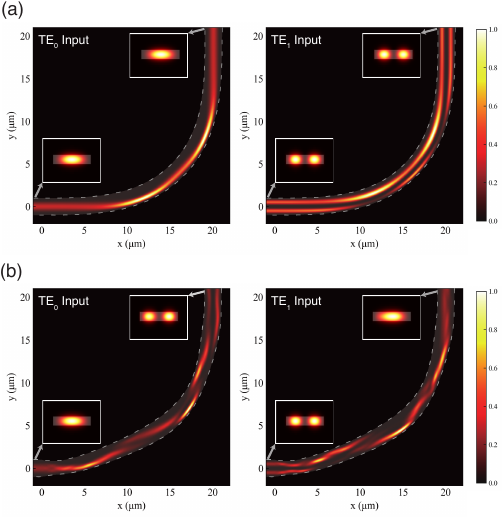}
\caption{Top-view power-flow and mode profiles for optimized bends with $R_{eff}=20\,\mu \mathrm{m}$ and input/output widths of $2\, \mu \mathrm{m}$.
(a) Mode-preserving bends.
(b) Mode-converting bends.
Insets show input and output mode profiles.}
\label{fig:reff20um_opt_topview}
\end{figure}

\begin{figure}[ht]
\centering
\includegraphics[width=0.5\linewidth]{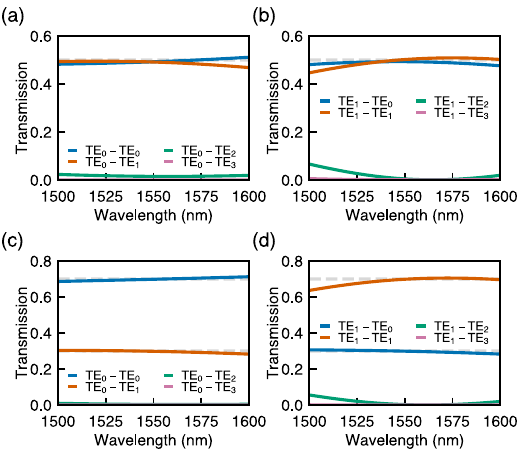}
\caption{Transmission spectra of optimized mode-splitting bends. (a), (b) 50:50 splitting for TE$_0$ and TE$_1$ input modes, respectively. (c), (d) 70:30 splitting for TE$_0$ and TE$_1$ inputs. Dashed lines indicate ideal transmission ratios.}
\label{fig:reff20um_opt_transmission}
\end{figure}

Additional optimization tasks for a larger bend with custom split ratios were performed similarly to demonstrate computational efficiency. For a mode-preserving bend, 6,000 simulations (100 particles × 60 iterations) took approximately 72 minutes using an Intel i9-9900K CPU. Optimizations of 50:50 mode-splitting bends required 8,000 simulations (100 particles × 80 iterations), completing in 68 minutes on an Intel i9-9900K CPU. A 70:30 mode-splitting bend, also involving 8,000 simulations (100 particles × 80 iterations), required 101 minutes using an Apple M2 CPU. More complex optimizations, such as mode-conversion bends, demanded greater computational resources, involving 16,000 simulations (80 particles × 200 iterations) and requiring approximately 425 minutes on an Apple M2 CPU.

Fig. \ref{fig:reff20um_opt_topview} illustrates the power flow profiles simulated by 3D FDTD for mode-preserving and mode-converting bends, respectively, demonstrating that each structure effectively achieves its intended design objectives. Fig. \ref{fig:reff20um_opt_transmission} presents optimized transmission results for mode-splitting bends obtained using 3D FDTD. For the 50:50 bend at the target wavelength (1550 nm), deviations from the target value at both outputs are below 1\%, with a bandwidth of 55 nm (1525-1580 nm) giving a deviation within 2\%. For the 70:30 bend, the deviations at 1550 nm from the target value are below 0.2\% at both outputs, with a bandwidth of 71 nm (1529–1600 nm) giving a deviation within 2\%. 

Note that the optimized structures are inherently adiabatic, which makes them compliant with standard foundry design rules and compatible with conventional fabrication processes. In addition, the designs exhibit robustness against fabrication-induced width errors on the order of tens of nanometers, as evidenced by our numerical analysis in Appendix \ref{app:optimized_structures}. Detailed descriptions of optimization procedures, optimized structural designs, and performance evaluations under various fabrication errors are provided in Appendix \ref{app:optimization}.

\section{Discussion}
We introduced a dataset-based simulation approach that leverages a precomputed geometric parameter space to significantly accelerate structural optimization compared to conventional electromagnetic field solvers. This framework enables rapid exploration of large design spaces, making the optimization of large-scale, multimode photonic circuits computationally feasible.

To illustrate the utility of this approach, we optimized waveguide bends using PSO, chosen for its wide acceptance and effectiveness in photonic optimization tasks. With a reasonably sized dataset (approximately 350 MB for rectangular waveguide structures), we achieved simulation times of less than one second per run, allowing optimization of waveguide bends for various design objectives in just a few hours. The methodology is also compatible with other design techniques\cite{liang_2021}, such as genetic algorithms\cite{GA_katoch2021}, CMA-ES\cite{Yu_2024}, and adjoint-enhanced hybrid methods\cite{Liao_2024_2}.

Although our demonstrations used rectangular waveguides due to their simplicity, the method can readily extend to other waveguide geometries. For instance, simulations involving different materials, such as silicon nitride (SiN), simply require the corresponding material-specific dataset. Likewise, more complex photonic structures, such as waveguide couplers, can be simulated by expanding the geometric parameter space to include additional parameters such as secondary waveguide widths or the gap between waveguides. Consequently, the applicability of this dataset-based approach is fundamentally constrained only by the theoretical limits of the EME itself.

Our method can also be applied to more complex structures, but this requires careful adjustment of dataset resolution to balance accuracy and computational cost. As the parameter space grows, generation and storage the dataset becomes more resource-intensive. This challenge can often be addressed by optimizing within limited parameter space, which is still sufficient to design many crucial components. For example, in case of direction coupler designs, constraints such as fixing the gap between waveguides or maintaining a constant combined width of top and bottom waveguides can be employed \cite{Cabanillas_2020, chen_2021}. In the future, techniques such as adaptive sampling or data compression may help improve scalability. However, simulating fully three-dimensional structures, such as grating couplers or metasurfaces, remains difficult and may require combining our approach with conventional EME or FDTD methods.

Our method's substantial reduction in computational demands creates new possibilities for exploiting multimode waveguide properties that were previously impractical to explore due to enormous computation resource requirement. Moreover, the approach integrates seamlessly into current photonic design workflows, providing a fast simulation back-end compatible with various optimization engines and layout tools. Coupled with existing foundry process design kits (PDKs) and automated layout generation tools, this technique accelerates design iterations, facilitating deeper exploration of photonic designs, and benefiting both academic research and commercial photonic circuit development.

\begin{appendix}
\section{Structure Generation}
\label{app:structure_generation}
Conventional simulation methods operate directly in real space, while dataset-based simulations inherently function within a geometric parameter space defined by datasets. Therefore, a systematic mapping between the real space and the geometric parameter space is essential for accurately simulating existing structures and generating novel designs.

\subsection{Conventional Structures}
To simulate conventional structures such as linear tapers, exponential tapers, arcs, Euler bends, or Bézier bends, a clear mapping from real space to geometric parameter space is necessary. This can be achieved by extracting geometric parameters along the propagation axis and treating them as a function of the propagation axis. For example, linear tapers can be described using linear width functions with zero curvature along the propagation axis. Similarly, curvature functions for bends can be implemented in a similar way. For more complex structures, whose width and curvature vary simultaneously, a more advanced function description is required. We chose an approach that first defines a center line and parameterizes the distance of the waveguide boundary from it. Such a method enables the extraction of both width and curvature as a function of the propagation axis. Once mapped, these continuous geometric parameters can be discretized according to the grid criteria of the predefined data set, including the grid spacing.

\subsection{Parameterized Random Adiabatic Structure Generation}

Generating arbitrary geometric structures through parameterization is essential, particularly for population-based optimization methods that rely on iterative random variations. Arbitrary taper designs require the definition of randomized width functions, whereas arbitrary bend structures utilize random curvature profiles along the propagation axis.

To ensure smooth geometric transitions, Bézier curves are employed to generate these arbitrary parameter profiles. A Bézier curve, parameterized by a variable $t \in [0,1]$, is defined as:
\begin{equation}
    \mathbf{B}(t) = 
    \begin{pmatrix} b_x(t) \\[4pt] b_y(t) \end{pmatrix}
    = \sum_{i=0}^{n} \frac{n!}{i!(n-i)!}(1 - t)^{n - i} t^i \mathbf{P}_i,
\end{equation}
where $\mathbf{P}_i$ are the control points and $n$ is a positive integer determining the complexity of the curve.

To parameterize a structure of known length $L$ with fixed input and output parameters $P_0$ and $P_n$, respectively, as well as predefined parameter bounds $P_{\text{min}}$ and $P_{\text{max}}$, the control points are defined as follows:

\begin{itemize}
    \item The endpoints are fixed as $\mathbf{P}_0 = (0, P_0)^T$ and $\mathbf{P}_n = (L, P_n)^T$, where $^T$ denotes transpose.
    \item Intermediate control points are determined by randomly selecting normalized values $P_i \in [0,1]$ and scaling according to:
    \begin{equation}
        \mathbf{P}_i = \left(\frac{i}{n}L,\;P_{\text{min}} + P_i(P_{\text{max}} - P_{\text{min}})\right)^T.
        \label{eq:bezier_control_points}
    \end{equation}
\end{itemize}

By construction, $b_x(t)$ is monotonically increasing, with a maximum value of $L$. Thus, the resulting parameter function $B(z)$ along the propagation axis length $z$ is defined by:
\begin{equation}
    B(z)\equiv b_y(b_x^{-1}(z)), \quad z\in [0,L].
    \label{eq:bezier_shape_func}
\end{equation}

Larger values of $n$ increase the flexibility and complexity of the resulting geometric structures. A schematic illustration of this parameterization approach is provided in the Fig. \ref{fig:reff10um_optimization}(a).

\section{Bend Parameterization}
\label{app:bend_parameterization}
Direct curvature parameterization can unpredictably alter bend size and angle. Specifically, generating the curvature function directly via the Bézier approach described earlier can result in changes to bend size and angle when parameters are varied. Consequently, a modified parameterization is necessary to satisfy constraints such as the effective radius ($R_{\text{eff}}$) and total bend angle ($\Theta$). This section describes a systematic method to construct a curvature function meeting these constraints, starting from a normalized parameter set $[P_0,P_1,\dots,P_n]$.

\subsection{Symmetric 90-Degree Bend}
We consider symmetric 90-degree bends characterized by an effective radius $R_{\mathrm{eff}}$. Without loss of generality, we assume the input port is located at the Cartesian coordinates $(0,0)$ with an initial angle of $0$, so the output port is at $(R_{\mathrm{eff}},R_{\mathrm{eff}})$ with an output angle of $\pi/2$ radians. The propagation distance from the input port is denoted as $z$.

The curvature $C(z)$ is related to the local bend radius $R(z)$ by $C(z)=1/R(z)$. For bends with nonnegative curvature, the local bend angle $\Theta(z)$ at propagation distance $z$ is obtained by:
\begin{equation}
    \Theta(z) = \int_0^z C(z')\,dz'.
\end{equation}

We define $z_{\pi/4}$ as the propagation length at which the bend angle reaches $\pi/4$ radians. Since $C(z)\geq 0$, this length is uniquely determined by:
\begin{equation}
    \int_0^{z_{\pi/4}} C(z')\,dz'=\frac{\pi}{4}.
    \label{eq:symmetric_point_condition}
\end{equation}

Due to symmetry, this point represents a reflection point along the bend, implying $C(z_{\pi/4}+z)=C(z_{\pi/4}-z)$. The Cartesian coordinates at this midpoint, $(x_{\pi/4}, y_{\pi/4})$, are determined by:
\begin{equation}
    x_{\pi/4}=\int_0^{z_{\pi/4}}\cos\Theta(z')\,dz', \quad y_{\pi/4}=\int_0^{z_{\pi/4}}\sin\Theta(z')\,dz',
\end{equation}
satisfying the constraint $R_{\mathrm{eff}}=x_{\pi/4}+y_{\pi/4}$.

To meet given constraints on bend angle and effective radius, we introduce a method for deriving curvature functions from an arbitrary nonnegative curvature shape. Consider a Bézier-generated function $B(z)$ obtained with normalized length $L=1$, and define a virtual curvature function $C_v(z)$ in the interval $[0,2]$ as:
\begin{equation}
    C_v(z) = 
    \begin{cases}
        B(z), & 0\leq z\leq 1,\\[4pt]
        B(2-z), & 1< z\leq 2.
    \end{cases}
\end{equation}

We generate a constraint-satisfying curvature function $C(z)$ by scaling and stretching the virtual function $C_v(z)$ as:
\begin{equation}
    C(z)=a\,C_v\left(\frac{z}{b}\right).
\end{equation}

Enforcing the bend-angle constraint from Eq.~(\ref{eq:symmetric_point_condition}) with $b=z_{\pi/4}$ yields a direct relation between parameters $a$ and $b$:
\begin{equation}
    b=z_{\pi/4}=\frac{\pi}{4a\int_0^1 C_v(z')\,dz'}.
    \label{eq:a_b_relation}
\end{equation}

Equation~(\ref{eq:a_b_relation}) indicates that decreasing $a$ increases $z_{\pi/4}$ and thus $R_{\mathrm{eff}}$, establishing a monotonic relationship between $R_{\mathrm{eff}}$ and parameter $a$. Consequently, for each desired effective radius, there exists exactly one set of parameters $(a, b)$ that simultaneously satisfies both the effective radius and total bend angle constraints.

\subsection{Asymmetric 90-degree Bend}

For asymmetric bends, the effective radii $R_x$ and $R_y$ along the $x$ and $y$ axes may differ, eliminating symmetry and imposing three independent constraints: the total bend angle, $R_x$, and $R_y$. To generate asymmetric bends that simultaneously satisfy these constraints, we employ the following two-step strategy:

\begin{enumerate}
  \item Generate an initial bend curvature that satisfies both the 90-degree bend angle and the $R_x$ constraint.
  \item Stretch the bend along the $y$-axis by a scaling factor $c_y$ to fulfill the $R_y$ constraint.
\end{enumerate}

In the asymmetric configuration, the radius $R_x$ is calculated by
\begin{equation}
    R_x = \int_0^{z_{\mathrm{max}}}\cos{\Theta(z')}\,dz'.
\end{equation}

In the first step, a virtual curvature function $C_v(z)$ is defined as:
\begin{equation}
    C_v(z) = B(z),\quad 0\leq z \leq 1,
\end{equation}
where $B(z)$ is generated using Eq.~(\ref{eq:bezier_shape_func}). Analogous to the symmetric bend approach, we construct the curvature function as $C(z)=a\,C_v(z/b)$. To enforce the bend-angle constraint, parameters $a$ and $b$ are related through:
\begin{equation}
    b = L = \frac{\pi}{2a\int_0^1 C_v(z)\,dz},
\end{equation}
where $L$ is the total propagation length. Thus, systematically varying $a$ yields a unique set $(a,b)$ that meets both the 90-degree angle and $R_x$ constraints.

In the second step, we apply a stretching transformation along the Cartesian $y$-axis by a factor $c_y$:
\begin{equation}
    (x(z), y(z)) \rightarrow (X(z), Y(z)) = (x(z), c_y y(z)).
\end{equation}

Consequently, the first and second derivatives become:
\begin{align}
    X'(z) &= x'(z) = \cos\Theta(z), \\
    Y'(z) &= c_y y'(z) = c_y\sin\Theta(z), \\
    X''(z) &= -\sin\Theta(z)\,C(z), \\
    Y''(z) &= c_y\cos\Theta(z)\,C(z).
\end{align}

Given the curvature definition
\begin{equation}
    C_{\mathrm{new}} = \frac{\left|X'Y'' - X''Y'\right|}{\left((X')^2 + (Y')^2\right)^{3/2}},
\end{equation}
the transformed curvature after stretching is expressed as:
\begin{equation}
    C_{\mathrm{new}}(z) = \frac{c_y\,C(z)}{\left(\cos^2\Theta(z)+c_y^2\sin^2\Theta(z)\right)^{3/2}}.
    \label{eq:new_curvature}
\end{equation}

The adjusted propagation length $z_{\mathrm{new}}$ is recalculated by:
\begin{equation}
    z_{\mathrm{new}} = \int_0^z\sqrt{\cos^2\Theta(u)+c_y^2\sin^2\Theta(u)}\,du.
    \label{eq:new_length}
\end{equation}

Combining Eqs.~(\ref{eq:new_curvature}) and (\ref{eq:new_length}), we obtain the final curvature function $C_{\mathrm{new}}(z_{\mathrm{new}})$, satisfying all three constraints: total bend angle, $R_x$, and $R_y$.

\section{Accuracy Test}
\label{app:test_structures}

To identify the optimal grid spacing, single-parameter structures varying solely in width or curvature were compared with reference FDTD results. After confirming convergence for these single-parameter cases, multi-parameter variation cases were evaluated at multiple grid resolutions, while the finest grid was used as a reference.

\subsection{Single-Parameter Test Structures}

Single-parameter test structures were generated systematically to evaluate the simulation accuracy. The width variation test sets were linear and Bézier-generated tapers, whereas the curvature variation test sets included partial Euler bends and Bézier-generated bends. The details of each test set are described below.

\textbf{Width variation test structures:}
Four subsets of tapers were constructed:
\begin{enumerate}
    \item Linear taper
    \item Bézier-shaped taper (4 control points): [1, 0, 1, 0]
    \item Bézier-shaped taper (12 control points): [1, 1, 0, 0, 1, 1, 0, 0, 1, 1, 0, 0]
    \item Bézier-shaped taper (10 control points): [0.4, 0.3, 0.5, 0.4, 0.7, 0.8, 0.4, 0.5, 0.5, 0.6]
\end{enumerate}
The control point tuples used in the Bézier-shaped tapers correspond to the set of $P_i$ terms defined in Eq.~(\ref{eq:bezier_control_points}), which define the width function. All taper structures had a fixed input width of $1~\mu\mathrm{m}$ and an output width of $3~\mu\mathrm{m}$, with zero curvature throughout their lengths. Each subset contained 10 structures, with total lengths ranging from $1~\mu\mathrm{m}$ to $10~\mu\mathrm{m}$ in steps of $1~\mu\mathrm{m}$.

\begin{figure}[htbp]
\centering
\includegraphics[width=.8\linewidth]{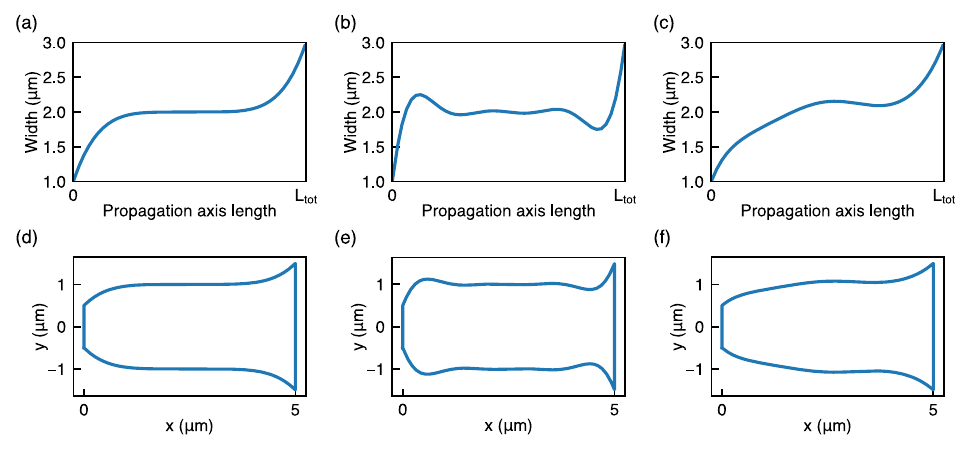}
\caption{Schematic of the Bézier shaped width variation test set.
(a)–(c) Width profiles generated using (a) 4-point control, (b) 12-point control, and (c) 10-point control along the propagation axis of length $L_{\mathrm{tot}}$. 
(d)–(f) Corresponding top-view taper shapes reconstructed from the width functions for a taper length of $5~\mu\mathrm{m}$, using (d) 4-point, (e) 12-point, and (f) 10-point control.}
\label{fig:Taper_test_set}
\end{figure}

Representative examples of width variation test structures are shown in Fig.~\ref{fig:Taper_test_set}. Width profiles generated with 4-, 12-, and 10-point control schemes are depicted in Fig.~\ref{fig:Taper_test_set}(a)–(c), respectively. The corresponding reconstructed top-view shapes for a representative taper length of $5~\mu\mathrm{m}$ are presented in Fig.~\ref{fig:Taper_test_set}(d)–(f), illustrating the geometric diversity used to evaluate width-dependent device performance.

\textbf{Curvature variation test structures:}
Curvature variation test structures consisted of partial Euler bends and Bézier-generated bends, all featuring a constant waveguide width of $2~\mu\mathrm{m}$ and a $90^\circ$ bend angle. The test set comprised four subsets (56 structures in total), each containing 14 structures with effective radii ranging from $10~\mu\mathrm{m}$ to $36~\mu\mathrm{m}$. The control point tuples used in the Bézier-shaped curvature correspond to the set of $P_i$ terms defined in Eq.~(\ref{eq:bezier_control_points}) and are employed to generate curvature profiles that satisfy constraints described in the bend parameterization section:
\begin{enumerate}
    \item Partial Euler bend (spiral portion, $p = 0.5$)
    \item Bézier-shaped curvature (4 control points): [1, 0, 1, 0]
    \item Bézier-shaped curvature (12 control points): [1, 1, 0, 0, 1, 1, 0, 0, 1, 1, 0, 0]
    \item Bézier-shaped curvature (10 control points): [0.4, 0.3, 0.5, 0.4, 0.7, 0.8, 0.4, 0.5, 0.5, 0.6]
\end{enumerate}

\begin{figure}[htbp]
\centering
\includegraphics[width=.8\linewidth]{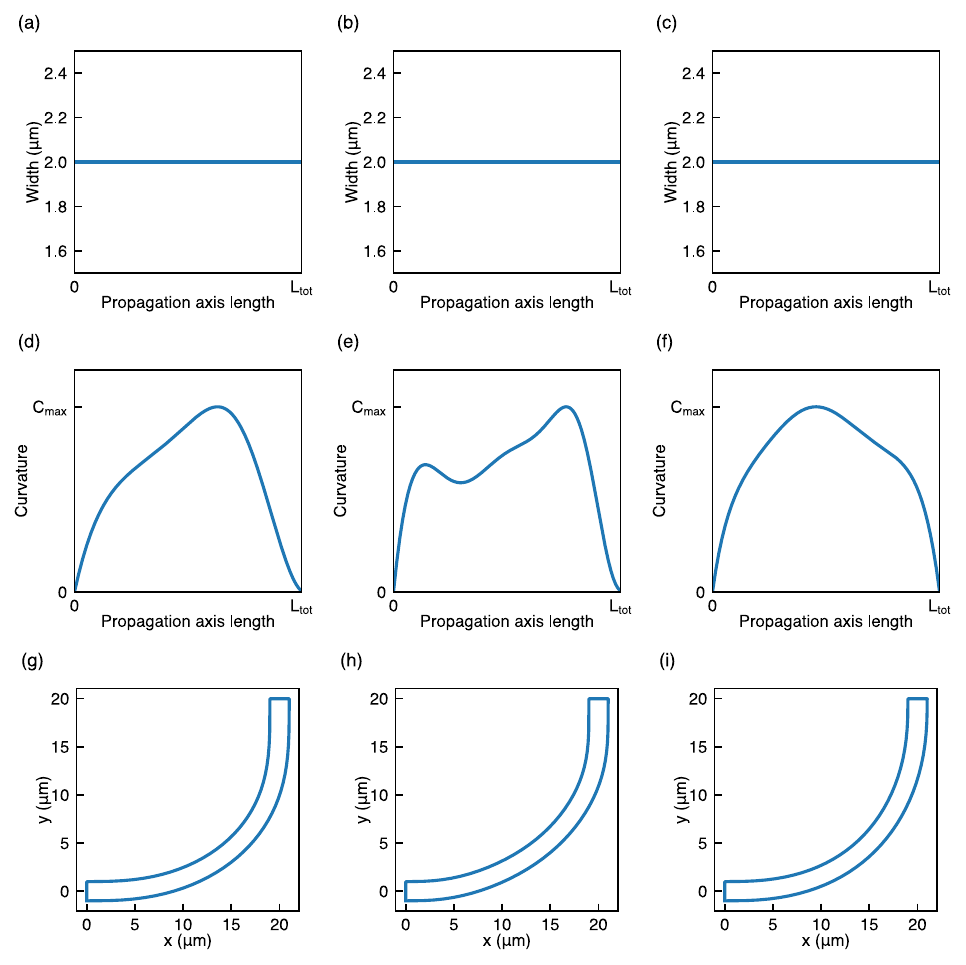}
\caption{Schematic of the Bézier shaped curvature variation test set.
(a)–(c) Width profiles generated using (a) 4-point control, (b) 12-point control, and (c) 10-point control along the propagation axis of length $L_{\mathrm{tot}}$. 
(d)–(f) Curvature profiles generated using (d) 4-point control, (e) 12-point control, and (f) 10-point control along the propagation axis of length $L_{\mathrm{tot}}$. 
(g)–(i) Corresponding top-view 90 degree bend shapes reconstructed from the width and curvature functions for a effective radius of $20~\mu\mathrm{m}$, using (g) 4-point, (h) 12-point, and (i) 10-point control.}
\label{fig:Curve_test_set}
\end{figure}

Representative examples of curvature variation structures are shown in Fig.~\ref{fig:Curve_test_set}. Width profiles (fixed at $2~\mu\mathrm{m}$) are presented in Fig.~\ref{fig:Curve_test_set}(a)–(c). Corresponding curvature profiles generated by 4-, 12-, and 10-point control schemes are illustrated in Fig.~\ref{fig:Curve_test_set}(d)–(f). Reconstructed 90-degree bend shapes for a representative effective radius of $20~\mu\mathrm{m}$ are depicted in Fig.~\ref{fig:Curve_test_set}(g)–(i). These examples highlight the geometric diversity resulting from varying control points and illustrate the range of curvature variations evaluated in the simulations.

\subsection{Multi-Parameter Test Structures}

A total of 93 structures were used to evaluate multiparameter datasets. All structures were configured as $90^\circ$ bends with equal effective radii in both axes ($R_x = R_y = R_{\mathrm{eff}}$) and waveguide widths fixed at $2~\mu\mathrm{m}$ at both input and output ports. The dataset was divided into three subsets based on the number of Bézier control points defining curvature and width functions. The control point tuples used in the Bézier-shaped width and curvature functions correspond to the $P_i$ terms defined in Eq.~(\ref{eq:bezier_control_points}), which define the width and curvature profiles, respectively:

\begin{enumerate}
    \item 2 control points: curvature control $[0.2, 0.8]$; width control $[0.6, 0.2]$
    \item 4 control points: curvature control $[0.9, 0.6, 0.1, 0.7]$; width control $[0.2, 0.7, 0.5, 0.7]$
    \item 10 control points: curvature control $[0.2, 0.8, 0.9, 0.2, 0.5, 0.3, 0.2, 1.0, 1.0, 0.8]$; width control $[0.4, 0.5, 0.5, 0.9, 0.4, 0.1, 0.5, 0.8, 0.5, 0.3]$
\end{enumerate}

Each subset contained 31 structures, with effective radii ($R_{\mathrm{eff}}$) ranging from $10~\mu\mathrm{m}$ to $40~\mu\mathrm{m}$ in increments of $1~\mu\mathrm{m}$.

\begin{figure}[htbp]
\centering
\includegraphics[width=.8\linewidth]{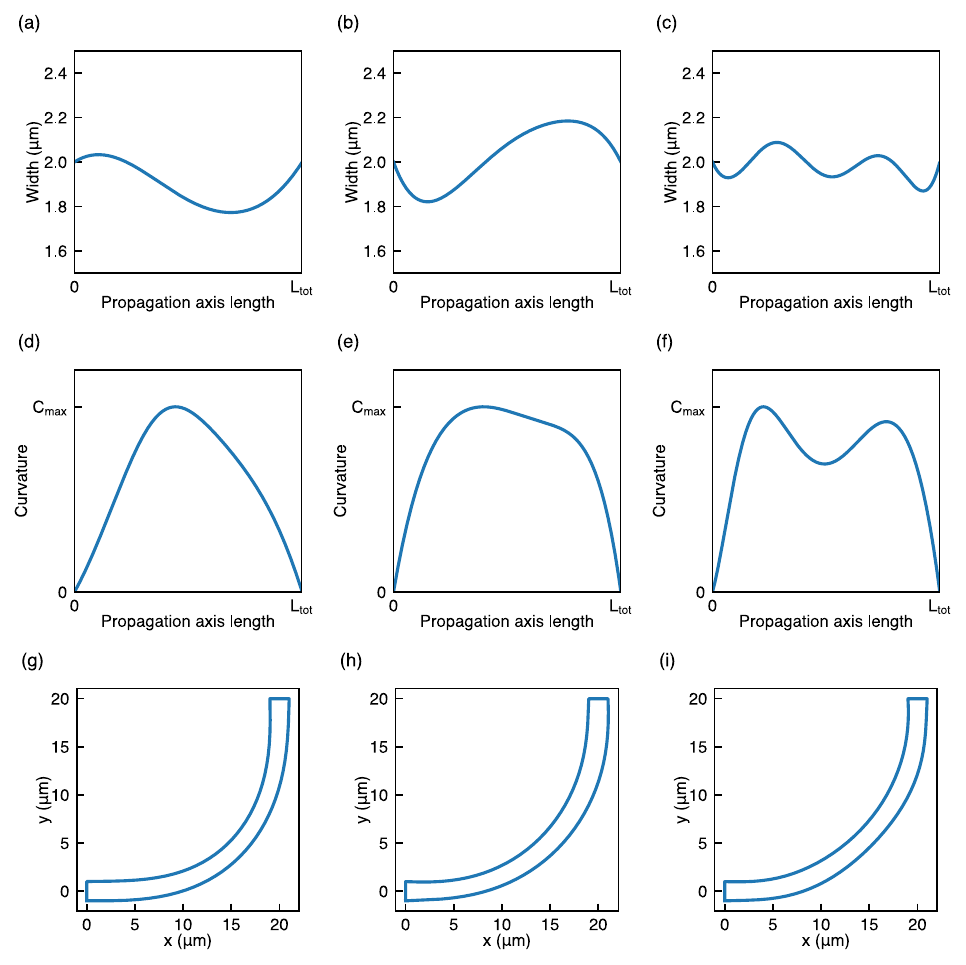}
\caption{Schematic of the multi-parameter test set.
(a)–(c) Width profiles generated using (a) 2-point control, (b) 4-point control, and (c) 10-point control along the propagation axis of length $L_{\mathrm{tot}}$. 
(d)–(f) Curvature profiles generated using (d) 2-point control, (e) 4-point control, and (f) 10-point control along the propagation axis of length $L_{\mathrm{tot}}$. 
(g)–(i) Corresponding top-view 90 degree bend shapes reconstructed from the width and curvature functions for a effective radius of $20~\mu\mathrm{m}$, using (g) 2-point, (h) 4-point, and (i) 10-point control.}
\label{fig:multiparam_test_set}
\end{figure}

Representative examples of multiparameter test structures are shown in Fig.~\ref{fig:multiparam_test_set}. Width profiles generated by 2-, 4-, and 10-point Bézier control are depicted in Fig.~\ref{fig:multiparam_test_set}(a)–(c), while corresponding curvature profiles are presented in Fig.~\ref{fig:multiparam_test_set}(d)–(f). The resulting 90-degree bend shapes reconstructed from these combined profiles, at a representative effective radius of $20~\mu\mathrm{m}$, are illustrated in Fig.~\ref{fig:multiparam_test_set}(g)–(i). These examples highlight the increasing geometric complexity associated with higher numbers of control points.

\section{Optimization}
\label{app:optimization}
\subsection{Optimization Method: Particle Swarm Optimization}
We employ Particle Swarm Optimization (PSO)\cite{PSO_Gad_2022} to optimize various bend structures. PSO is a widely used metaheuristic optimization algorithm inspired by swarm intelligence (SI), particularly the collective behavior observed in bird flocking. The algorithm operates with a population of \(N\) particles, where each particle explores the \(D\)-dimensional parameter space to locate the optimal solution.

Each particle maintains a record of its own best-found position, while the swarm shares information regarding the globally best-known position. Based on these, the particle updates its velocity and position iteratively. For a minimization problem, the best-known position of the \(i\)-th particle at iteration \(t\) is denoted as:
\begin{equation}
    \mathbf{p}^t_{\text{best},i} = \mathbf{x}^*_i \quad \text{such that} \quad f(\mathbf{x}^*_i) = \min_{k=1,2,\dots,t} \{f(\mathbf{x}^k_i)\},
\end{equation}
where \(i \in \{1, 2, \dots, N\}\). 

The global best position across all particles up to iteration \(t\) is given by:
\begin{equation}
    \mathbf{g}^t_{\text{best}} = \mathbf{x}^t_* \quad \text{such that} \quad f(\mathbf{x}^t_*) = \min_{\substack{i=1,\dots,N \\ k=1,\dots,t}} \{f(\mathbf{x}^k_i)\}.
\end{equation}

In each iteration, the velocity of the \(i\)-th particle is updated as:
\begin{equation}
    \mathbf{v}^{t+1}_i = \omega \mathbf{v}^t_i + c_1 \mathbf{r}_1 \odot (\mathbf{p}^t_{\text{best},i} - \mathbf{x}^t_i) + c_2 \mathbf{r}_2 \odot (\mathbf{g}^t_{\text{best}} - \mathbf{x}^t_i),
\end{equation}
and the position is updated accordingly:
\begin{equation}
\mathbf{x}_i^{t+1}=\mathbf{x}_i^t+\mathbf{v}_i^{t+1}.
\end{equation}
where \(\omega\) is the inertia weight that controls the influence of a particle’s previous velocity on its current motion. The vectors \(\mathbf{r}_1\) and \(\mathbf{r}_2\) are randomly generated at each iteration, with each component uniformly distributed in the range \([0,1]^D\). The coefficients \(c_1\) and \(c_2\) are user-defined hyperparameters that balance the influence of individual (cognitive) and swarm (social) knowledge, respectively. To prevent overly large updates, the particle velocity is often clamped to a predefined maximum value.

For bend optimization, we employ a modified PSO approach in which the width and curvature parameters are treated independently. Each particle’s position vector \(\mathbf{x}_i\) is composed of two components—width \(\mathbf{w}_i\) and curvature \(\mathbf{c}_i\)—and is defined as:
\begin{equation}
\mathbf{x}_i = 
\begin{pmatrix}
\mathbf{w}_i\\
\mathbf{c}_i
\end{pmatrix}.
\end{equation}

Correspondingly, the individual best position \(\mathbf{p}^t_{\text{best},i}\) and global best position \(\mathbf{g}^t_{\text{best}}\) are represented as:
\begin{equation}
\mathbf{p}^t_{\text{best},i} = 
\begin{pmatrix}
\mathbf{p}^t_{\mathbf{w},\text{best},i}\\
\mathbf{p}^t_{\mathbf{c},\text{best},i}
\end{pmatrix}, \quad
\mathbf{g}^t_{\text{best}} = 
\begin{pmatrix}
\mathbf{g}^t_{\mathbf{w},\text{best}}\\
\mathbf{g}^t_{\mathbf{c},\text{best}}
\end{pmatrix}.
\end{equation}

The velocity updates for the width and curvature components are performed separately. For the width component:
\begin{equation}
\mathbf{v}^{t+1}_{\mathbf{w},i} = \omega_{\mathbf{w}} \mathbf{v}^t_{\mathbf{w},i} + c_{\mathbf{w},1} \mathbf{r}_{\mathbf{w},1} \odot (\mathbf{p}^t_{\mathbf{w},\text{best},i} - \mathbf{w}^t_i) + c_{\mathbf{w},2} \mathbf{r}_{\mathbf{w},2} \odot (\mathbf{g}^t_{\mathbf{w},\text{best}} - \mathbf{w}^t_i),
\end{equation}
and for the curvature component:
\begin{equation}
\mathbf{v}^{t+1}_{\mathbf{c},i} = \omega_{\mathbf{c}} \mathbf{v}^t_{\mathbf{c},i} + c_{\mathbf{c},1} \mathbf{r}_{\mathbf{c},1} \odot (\mathbf{p}^t_{\mathbf{c},\text{best},i} - \mathbf{c}^t_i) + c_{\mathbf{c},2} \mathbf{r}_{\mathbf{c},2} \odot (\mathbf{g}^t_{\mathbf{c},\text{best}} - \mathbf{c}^t_i).
\end{equation}

\FloatBarrier
\subsection{Optimized Structures}
\label{app:optimized_structures}

In this section we describe the objective function used in each optimization of the structure and the detailed structural parameter function and top-view shape. Also, we show the detailed verified performances of transmission or insertion loss in the range from 1500 nm to 1600 nm wavelength. All verifications are performed using 3D FDTD with mesh size $\lambda / 22$.

\subsubsection{Mode-Preserving Bend with Effective Radius of $10~\mu\textrm{m}$}
A symmetric bend was designed to implement the mode-preserving function for multi-mode waveguides with an effective radius of $10~\mu\mathrm{m}$. The FOM history, optimized parameter functions, and reconstructed bend shape are shown in the main text, Fig. \ref{fig:reff10um_optimization}(b)–(d).

\begin{figure}[htbp]
\centering
\includegraphics[width=.8\linewidth]{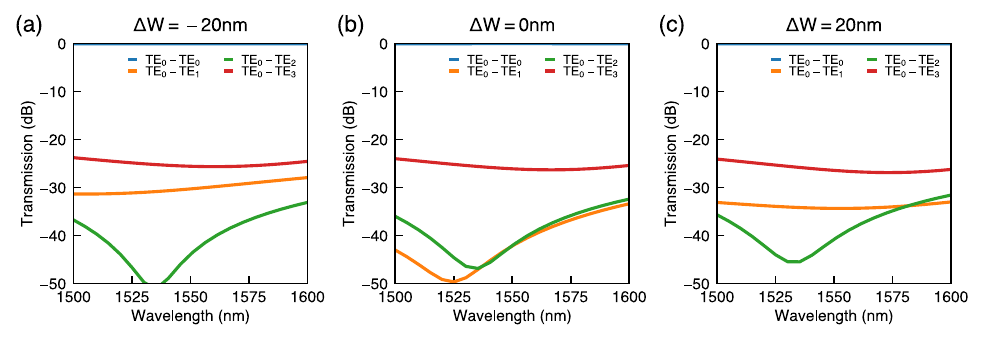}
\caption{Transmission spectra of a mode-preserving bend with a $10~\mu\textrm{m}$ radius for TE$_0$ input under width deviations of (a) -20 nm, (b) 0 nm, and (c) +20 nm. Crosstalk remains suppressed despite moderate fabrication errors.}
\label{fig:reff10um_transmission}
\end{figure}

Figure~\ref{fig:reff10um_transmission} presents the transmission spectra for TE$_0$ input under width deviations of $\Delta W = -20$~nm, $0$~nm, and $+20$~nm. This study provides information on structures robustness against common fabrication errors. In all cases, the power remains predominantly in the TE$_0$ mode, and intermodal crosstalk is effectively suppressed over the 1500–1600~nm wavelength range. 

\begin{figure}[htbp]
\centering
\includegraphics[width=.6\linewidth]{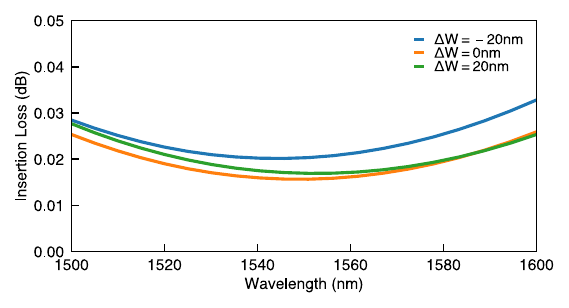}
\caption{Insertion loss for a mode-preserving bend with an effective radius of $10\,\mu\mathrm{m}$ under different width fabrication error conditions. In all cases, the insertion loss remains below 0.035 dB across the 1500–1600 nm, demonstrating high robustness to fabrication-induced width variations.}
\label{fig:mode_preserving_10um_bend_insertion_loss}
\end{figure}

The corresponding insertion loss is shown in Fig.~\ref{fig:mode_preserving_10um_bend_insertion_loss}. For all structures tested with different width errors, the insertion loss remains below 0.035~dB across the entire wavelength range. This confirms that the structure maintains low-loss performance and high fabrication tolerance at reduced bend radii.

\FloatBarrier
\subsubsection{Mode-Preserving Bend with Effective Radius of $20~\mu\textrm{m}$}

The figure of merit (FOM) is defined as
\begin{equation}
    \mathrm{FOM} = \left| 3 - |S_{11}|^2 - |S_{22}|^2 - |S_{33}|^2 \right|,
\end{equation}
and is minimized when the input mode is fully preserved across all ports.

\begin{figure}[htbp]
\centering
\includegraphics[width=.6\linewidth]{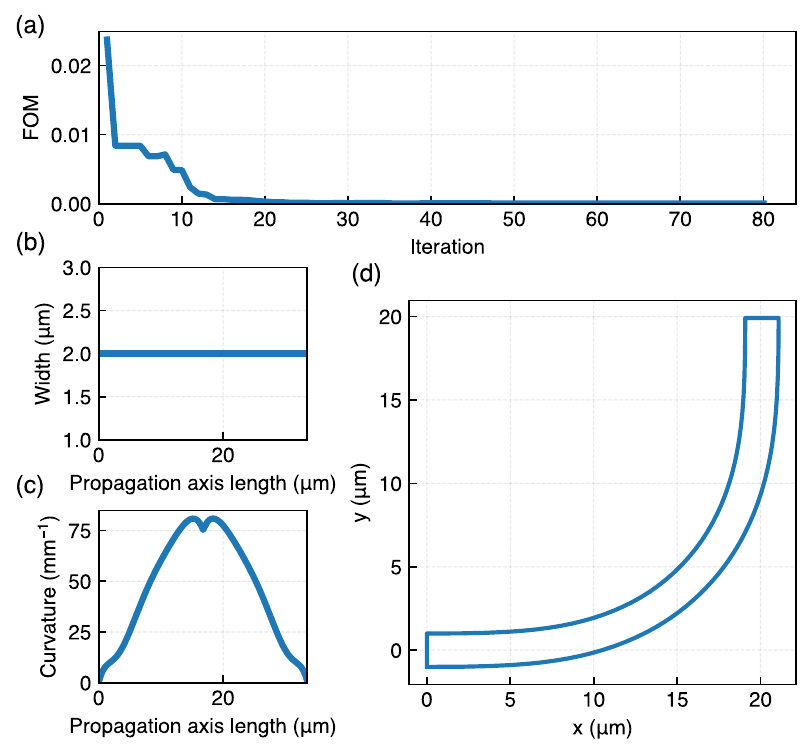}
\caption{Optimization results for a mode-preserving bend with an effective radius of $20\,\mu\mathrm{m}$. 
(a) Evolution of the figure of merit (FOM) over optimization iterations. 
(b), (c) Optimized geometric parameter functions for (b) width and (c) curvature along the propagation axis. 
(d) Top-view of the resulting bend shape reconstructed from the optimized width and curvature profiles.
}
\label{fig:mode_preserving_20um_bend}
\end{figure}

A symmetric mode-preserving bend was designed with an effective radius of $20~\mu\mathrm{m}$. Since the conventional partial Euler bend already provides good mode preservation with constant width, we fixed the waveguide width and focused the optimization solely on the curvature profile to maintain adiabaticity. The optimization history is shown in Fig.~\ref{fig:mode_preserving_20um_bend}(a), indicating rapid convergence. The optimized width and curvature profiles are provided in Fig.~\ref{fig:mode_preserving_20um_bend}(b) and (c), respectively, and the reconstructed bend shape is shown in Fig.~\ref{fig:mode_preserving_20um_bend}(d).

\begin{figure}[htbp]
\centering
\includegraphics[width=.8\linewidth]{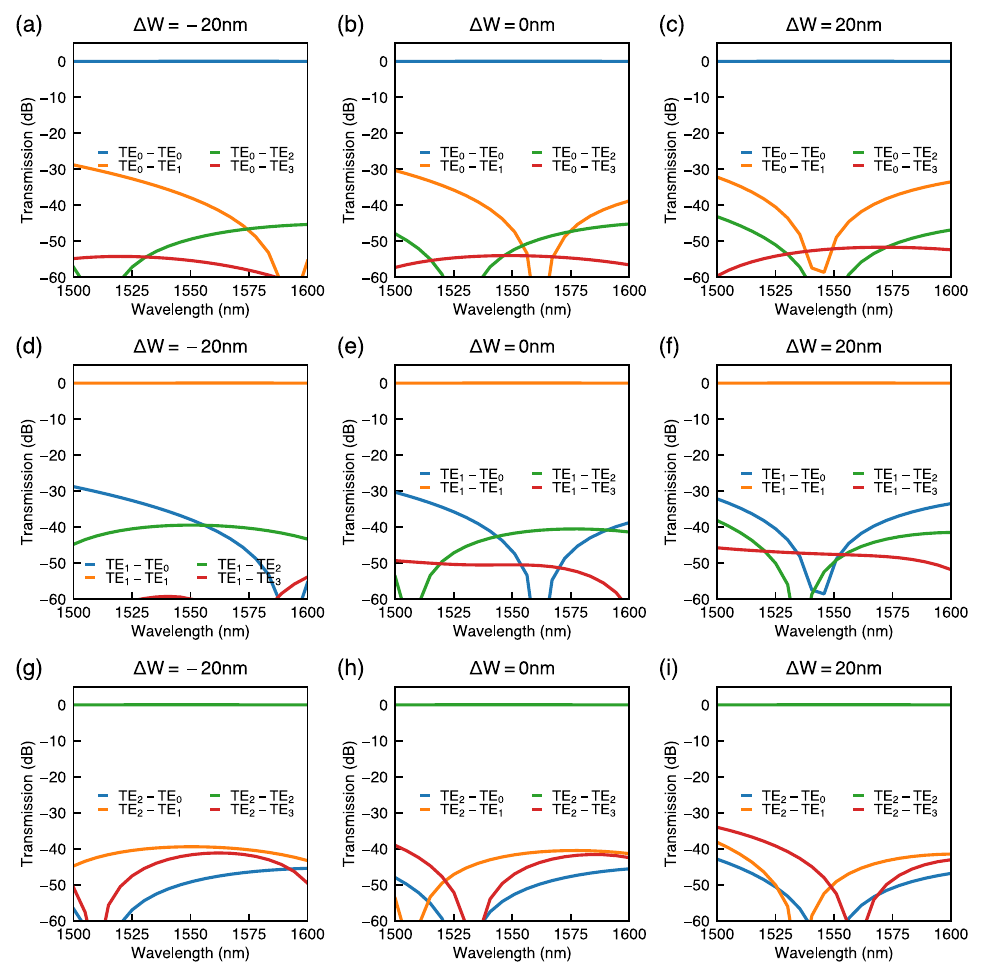}
\caption{Transmission spectra for a mode-preserving bend with an effective radius of $20\,\mu\mathrm{m}$ under different width fabrication error conditions. 
(a)–(c) $\mathrm{TE}_0$ input, (d)–(f) $\mathrm{TE}_1$ input, and (g)–(i) $\mathrm{TE}_2$ input. 
Left column (a, d, g): transmission with a width deviation of $-20\,\mathrm{nm}$; 
center column (b, e, h): no width deviation ($\Delta W = 0$); 
right column (c, f, i): transmission with a width deviation of $+20\,\mathrm{nm}$. 
The results confirm that intermodal crosstalk remains well suppressed even in the presence of realistic width fabrication errors.
}
\label{fig:mode_preserving_20um_bend_transmission}
\end{figure}

\begin{figure}[htbp]
\centering
\includegraphics[width=.8\linewidth]{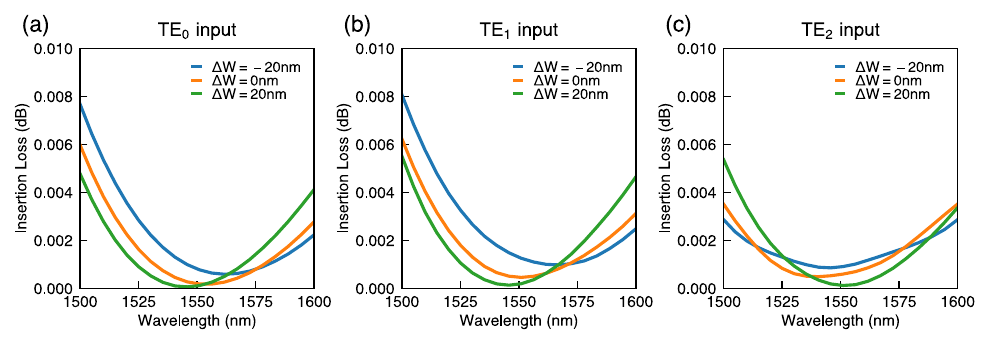}
\caption{Insertion loss for a mode-preserving bend with an effective radius of $20\,\mu\mathrm{m}$ under different width fabrication error conditions. 
(a) $\mathrm{TE}_0$ input, (b) $\mathrm{TE}_1$ input, and (c) $\mathrm{TE}_2$ input. 
The three traces in each plot represent cases with $\Delta W = -20\,\mathrm{nm}$ (blue), $\Delta W = 0$ (orange), and $\Delta W = +20\,\mathrm{nm}$ (green), respectively. 
In all cases, the insertion loss remains below 0.01 dB across the 1500–1600 nm, demonstrating high robustness to fabrication-induced width variations.}
\label{fig:mode_preserving_20um_bend_insertion_loss}
\end{figure}

Figure~\ref{fig:mode_preserving_20um_bend_transmission} presents the transmission spectra for multiple input modes under width deviations of $\Delta W = \pm 20$~nm. The design maintains robust mode preservation with suppressed intermodal crosstalk across the 1500–1600~nm range. As shown in Fig.~\ref{fig:mode_preserving_20um_bend_insertion_loss}, insertion loss remains below 0.01~dB under all tested conditions, confirming high tolerance to fabrication-induced width variations.

\FloatBarrier
\subsubsection{Mode-converting Bend with Effective Radius of $20~\mu\textrm{m}$}
The figure of merit (FOM) is defined as
\begin{equation}
    \mathrm{FOM} = 2 - \left( |S_{12}|^2 + |S_{21}|^2 \right),
\end{equation}
and is minimized when full power transfer occurs between the input and target output modes, indicating ideal mode conversion.

\begin{figure}[htbp]
\centering
\includegraphics[width=.6\linewidth]{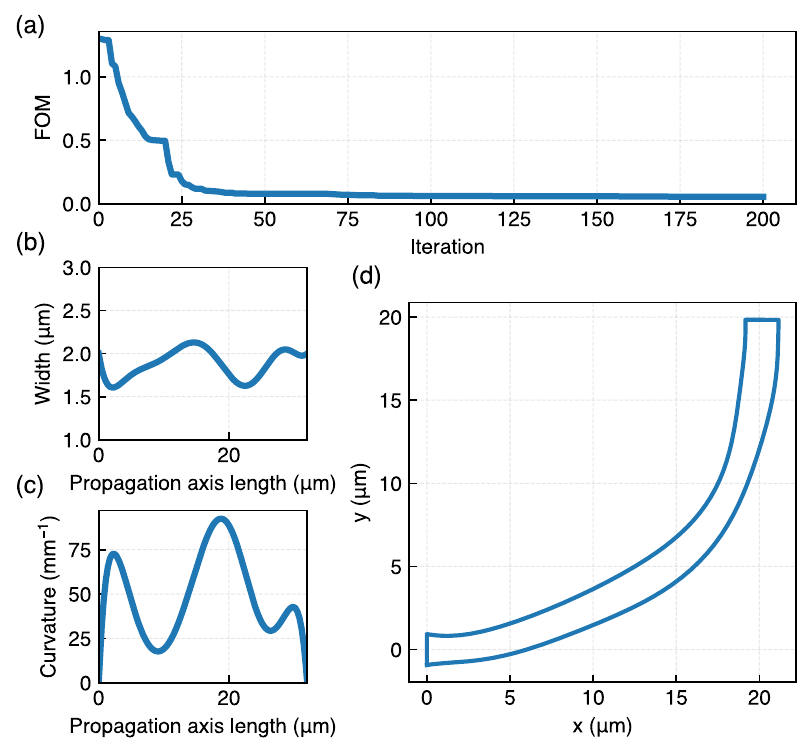}
\caption{Optimization results for a mode-converting bend with an effective radius of $20\,\mu\mathrm{m}$. 
(a) Evolution of the figure of merit (FOM) over optimization iterations. 
(b), (c) Optimized geometric parameter functions for (b) width and (c) curvature along the propagation axis. 
(d) Top-view of the resulting bend shape reconstructed from the optimized width and curvature profiles.}
\label{fig:mode_converting_bend}
\end{figure}

An asymmetric bend with an effective radius of $20~\mu\mathrm{m}$ was designed to enable efficient mode conversion. The optimization history is shown in Fig.~\ref{fig:mode_converting_bend}(a), which shows rapid and stable convergence. The optimized width and curvature profiles are presented in Fig.~\ref{fig:mode_converting_bend}(b) and (c), and the corresponding bend shape reconstructed from these parameters is shown in Fig.~\ref{fig:mode_converting_bend}(d).

\begin{figure}[htbp]
\centering
\includegraphics[width=.8\linewidth]{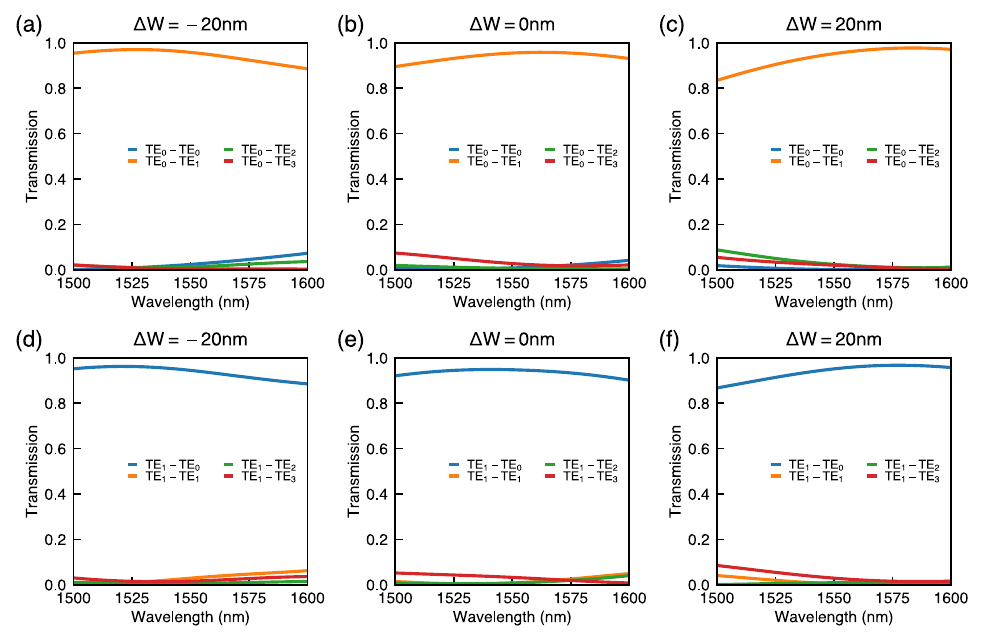}
\caption{Transmission spectra for a mode-converting bend with an effective radius of $20\,\mu\mathrm{m}$ under different width fabrication error conditions. 
(a)–(c) show results for $\mathrm{TE}_0$ input, and (d)–(f) for $\mathrm{TE}_1$ input. 
Left column: $\Delta W = -20\,\mathrm{nm}$; 
center column: $\Delta W = 0\,\mathrm{nm}$; 
right column: $\Delta W = +20\,\mathrm{nm}$. 
The intended mode conversion remains dominant across the full 1500–1600 nm wavelength range.}
\label{fig:mode_converting_20um_bend_transmission}
\end{figure}

Figure~\ref{fig:mode_converting_20um_bend_transmission} shows the transmission spectra for TE$_0$ and TE$_1$ inputs across the 1500–1600~nm wavelength range under width deviations of $\Delta W = \pm 20$~nm and 0~nm. In all cases, the intended mode conversion remains dominant, with minimal crosstalk to undesired modes. The results confirm the robustness of the design to fabrication-induced width variations.

\FloatBarrier
\subsubsection{50:50 Mode-splitting Bend with Effective Radius of $20~\mu\textrm{m}$}
The figure of merit (FOM) is defined as
\begin{equation}
    \mathrm{FOM} = \left|0.5 - |S_{11}|\right|^2 + \left|0.5 - |S_{12}|\right|^2 + \left|0.5 - |S_{21}|\right|^2 + \left|0.5 - |S_{22}|\right|^2,
\end{equation}
and is minimized when equal power is split between the TE$_0$ and TE$_1$ modes at the output, regardless of whether the input is TE$_0$ or TE$_1$.

\begin{figure}[htbp]
\centering
\includegraphics[width=.6\linewidth]{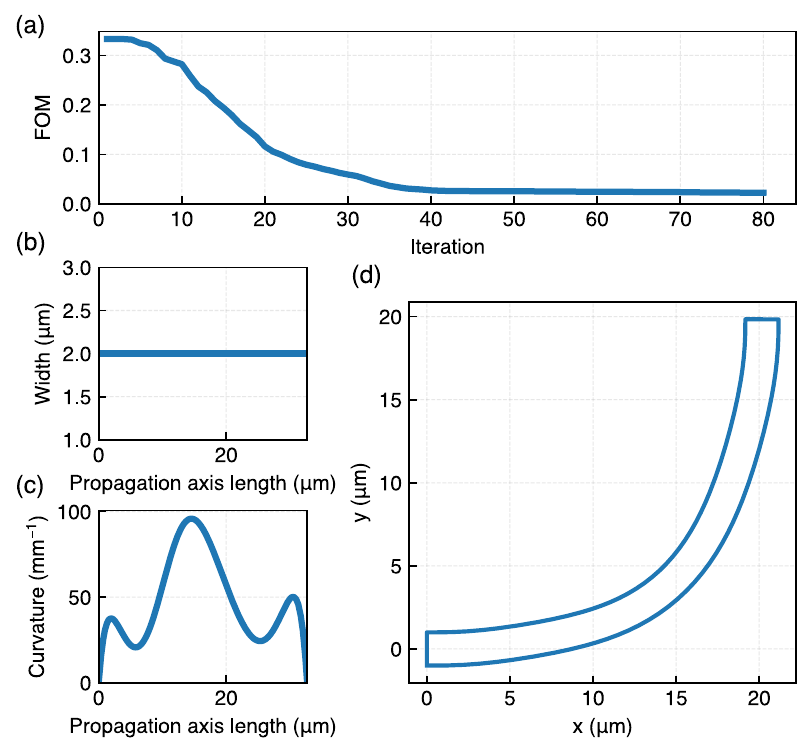}
\caption{Optimization results for a 50:50 mode-splitting bend with an effective radius of $20\,\mu\mathrm{m}$. 
(a) Evolution of the figure of merit (FOM) over optimization iterations. 
(b), (c) Optimized geometric parameter functions for (b) width and (c) curvature along the propagation axis. 
(d) Top-view of the resulting bend shape reconstructed from the optimized width and curvature profiles.}
\label{fig:mode_splitting_bend_5050}
\end{figure}

An asymmetric bend with an effective radius of $20~\mu\mathrm{m}$ was used to implement the 50:50 mode-splitting function. The waveguide width was fixed at $2~\mu\mathrm{m}$ to simplify the design. The optimization history in Fig.~\ref{fig:mode_splitting_bend_5050}(a) shows a smooth convergence towards the target splitting condition. The optimized width and curvature profiles, along with the corresponding bend shape, are shown in Fig.~\ref{fig:mode_splitting_bend_5050}(b)–(d).

\begin{figure}[htbp]
\centering
\includegraphics[width=.8\linewidth]{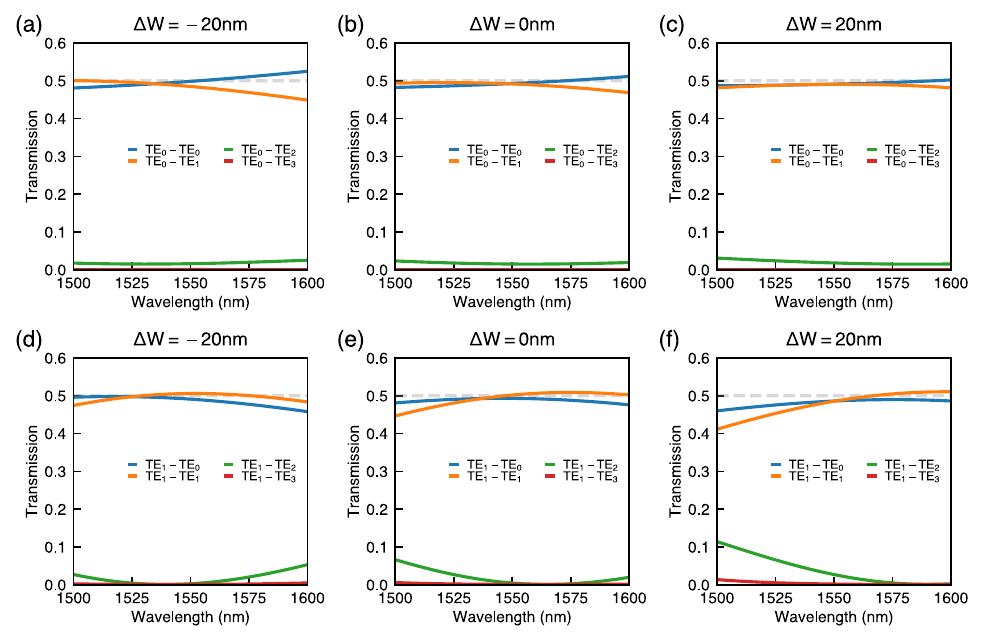}
\caption{Transmission spectra of a 50:50 mode-splitting bend with an effective radius of $20\,\mu\mathrm{m}$ under various width fabrication error conditions. 
(a)–(c) show results for $\mathrm{TE}_0$ input, and (d)–(f) for $\mathrm{TE}_1$ input. 
Left column: $\Delta W = -20\,\mathrm{nm}$; 
center column: $\Delta W = 0\,\mathrm{nm}$; 
right column: $\Delta W = +20\,\mathrm{nm}$. 
The designed 50:50 mode splitting is maintained across the 1500–1600\,nm wavelength range. The gray dashed line represents the target transmission level defined during optimization.}
\label{fig:5050_splitting_bend_transmission}
\end{figure}

Figure~\ref{fig:5050_splitting_bend_transmission} presents the transmission spectra under width deviations of $\Delta W = \pm20$~nm and 0~nm. The target transmission level of 0.5 is consistently maintained across the 1500–1600~nm wavelength range, as indicated by the gray dashed line. The design demonstrates high robustness to fabrication-induced width variations while preserving the desired 50:50 power splitting behavior.

\FloatBarrier
\subsubsection{70:30 Mode-splitting Bend with Effective Radius of $20~\mu\textrm{m}$}
The figure of merit (FOM) is defined as
\begin{equation}
    \mathrm{FOM} = \left|0.7 - |S_{11}|\right|^2 + \left|0.3 - |S_{12}|\right|^2 + \left|0.3 - |S_{21}|\right|^2 + \left|0.7 - |S_{22}|\right|^2,
\end{equation}
and is minimized when the transmission between ports achieves the target 70:30 power splitting ratio for both TE$_0$ and TE$_1$ inputs.

\begin{figure}[htbp]
\centering
\includegraphics[width=.6\linewidth]{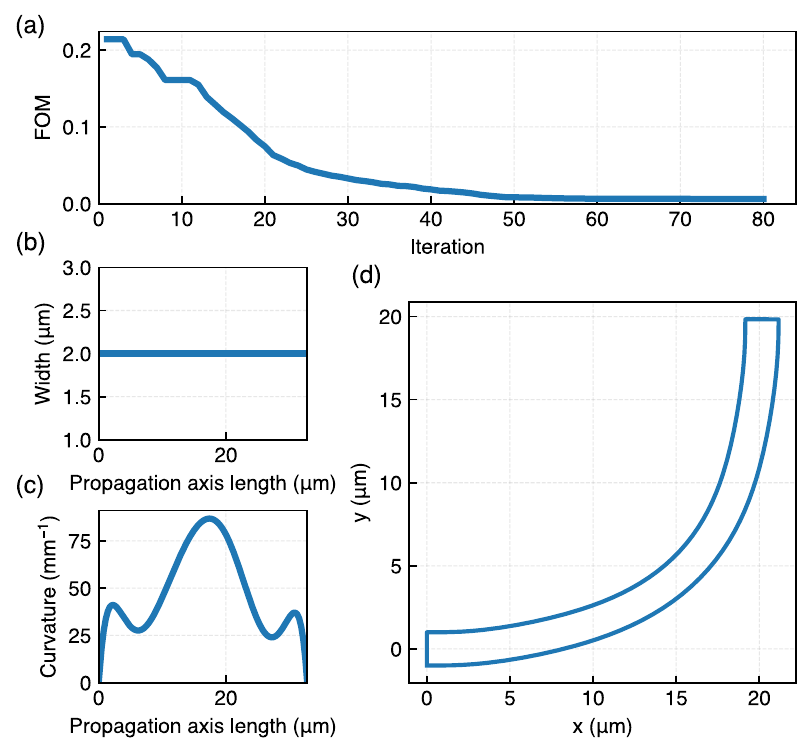}
\caption{Optimization results for a 70:30 mode-splitting bend with an effective radius of $20\,\mu\mathrm{m}$. 
(a) Evolution of the figure of merit (FOM) over optimization iterations. 
(b), (c) Optimized geometric parameter functions for (b) width and (c) curvature along the propagation axis. 
(d) Top-view of the resulting bend shape reconstructed from the optimized width and curvature profiles.}
\label{fig:mode_splitting_bend_7030}
\end{figure}

An asymmetric bend structure with an effective radius of $20~\mu\mathrm{m}$ was used to realize the desired mode-splitting behavior. The width of the waveguide was fixed at $2~\mu\mathrm{m}$ and focused the optimization solely on the curvature profile. The optimization history, shown in Fig.~\ref{fig:mode_splitting_bend_7030}(a), indicates smooth convergence. The optimized width and curvature profiles are presented in Fig.~\ref{fig:mode_splitting_bend_7030}(b) and (c), and the corresponding bend shape is shown in Fig.~\ref{fig:mode_splitting_bend_7030}(d).

\begin{figure}[htbp]
\centering
\includegraphics[width=.8\linewidth]{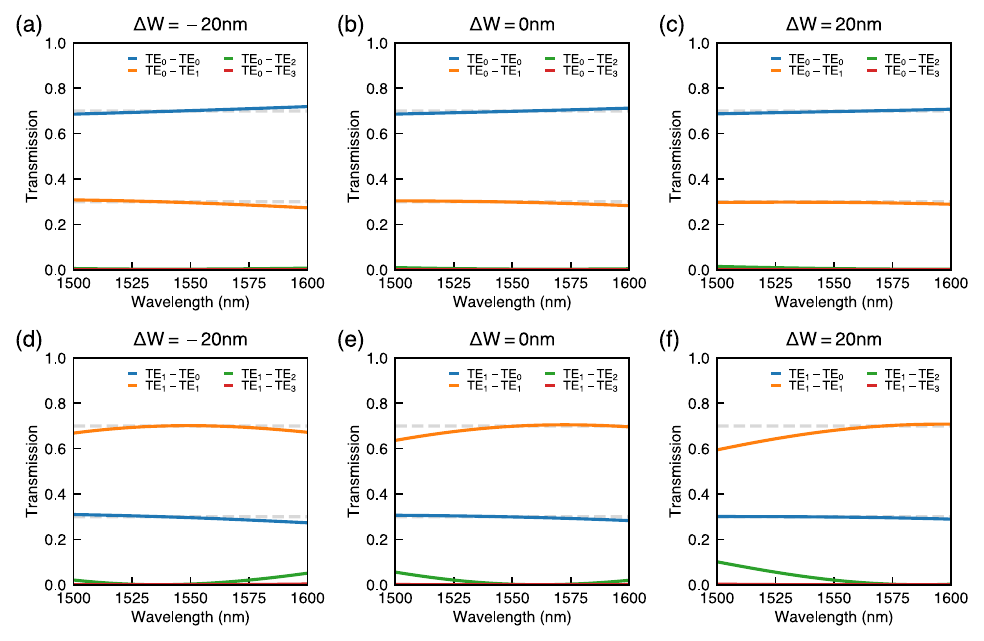}
\caption{Transmission spectra of a 70:30 mode-splitting bend with an effective radius of $20\,\mu\mathrm{m}$ under various width fabrication error conditions. 
(a)–(c) show results for $\mathrm{TE}_0$ input, and (d)–(f) for $\mathrm{TE}_1$ input. 
Left column: $\Delta W = -20\,\mathrm{nm}$; 
center column: $\Delta W = 0\,\mathrm{nm}$; 
right column: $\Delta W = +20\,\mathrm{nm}$. 
The designed 70:30 mode splitting is consistently maintained across the 1500–1600\,nm wavelength range, even in the presence of width variations. 
Gray dashed lines indicate the target transmission levels defined during optimization.}
\label{fig:7030_splitting_bend_transmission}
\end{figure}

Figure~\ref{fig:7030_splitting_bend_transmission} shows the transmission spectra under width deviations of $\Delta W = \pm 20$~nm and 0~nm. The target transmission levels of 0.7 and 0.3, marked by gray dashed lines, are consistently maintained across the 1500–1600~nm wavelength range. These results confirm robust 70:30 mode-splitting performance and strong tolerance to fabrication-induced width variations.

\FloatBarrier
\end{appendix}

\begin{backmatter}
\bmsection{Funding}
National Research Foundation of Korea (RS-2021-NR061364, RS-2022-NR068818, RS-2024-00442762, NRF-2022M3E4A1077013, NRF-2022M3H3A1085772); Institute of Informations \& Communications Technology Planning \& Evaluation (RS-2024-00397959)

\bmsection{Acknowledgment}

\bmsection{Disclosures} The authors declare no conflicts of interest.

\bmsection{Data availability} Data underlying the results presented in this paper are not publicly available at this time but may be obtained from the authors upon reasonable request.

\bigskip

\end{backmatter}


\bibliography{main}

\end{document}